\documentclass[11pt]{article}
\usepackage{amsmath,amssymb,amsfonts,amsthm,thmtools,bm,bbm,mathrsfs,natbib,tocloft,comment,accents,geometry,enumitem}
\usepackage[onehalfspacing]{setspace}
\usepackage{cancel,sectsty,color,array,hyperref,graphicx}

\hypersetup{pdfborder = {0 0 0},colorlinks=true,linkcolor=blue,urlcolor=blue,citecolor=blue}
\newtheorem{theorem}{Theorem}
\newtheorem{lemma}{Lemma}
\newtheorem{assumption}{Assumption}

\newenvironment{myproof}[1] {\noindent\textbf{Proof of {#1}.}}{\hfill$\square$\bigskip}

\newenvironment{examp}
{
    \bigskip\par\noindent\textbf{Example}.
}
{
    \hfill{\LARGE$\lrcorner$}\medskip
}

\newcommand{\I}{\mathbbm{1}}

\DeclareMathOperator*{\argmax}{arg\,max}

\DeclareMathOperator*{\sgn}{sgn}

\newcommand{\E}{\mathbb{E}}

\newcommand{\N}{\mathbb{N}}
\renewcommand{\P}{\mathbb{P}}
\newcommand{\R}{\mathbb{R}}
\renewcommand{\S}{\mathbb{S}}
\newcommand{\V}{\mathbb{V}}

\newcommand{\Be}{\mathbf{e}}

\newcommand{\Bh}{\mathbf{h}}

\newcommand{\Bs}{\mathbf{s}}
\newcommand{\Bt}{\mathbf{t}}

\newcommand{\Bw}{\mathbf{w}}
\newcommand{\Bx}{\mathbf{x}}

\newcommand{\Bz}{\mathbf{z}}

\newcommand{\Bbeta}{\bm{\beta}}

\newcommand{\Bdelta}{\bm{\delta}}

\newcommand{\Btheta}{\bm{\theta}}

\newcommand{\Blambda}{\bm{\lambda}}

\newcommand{\Bomega}{\bm{\omega}}

\newcommand{\CB}{\mathscr{B}}
\newcommand{\CC}{\mathcal{C}}

\newcommand{\CG}{\mathcal{G}}

\newcommand{\CN}{\mathcal{N}}

\newcommand{\CX}{\mathcal{X}}

\allowdisplaybreaks

\begin{document}

\title{Continuity of the Distribution Function of the $\argmax$ of a Gaussian Process\thanks{We are grateful to Boris Hanin, Jason Klusowski, Ulrich M\"uller, Mathieu Rosenbaum, Misha Shkolnikov, Ronnie Sircar, Will Underwood, and an anonymous referee for insightful discussions and comments. Cattaneo gratefully acknowledges financial support from the National Science Foundation through grants SES-1947805, DMS-2210561, and SES-2241575, and Jansson gratefully acknowledges financial support from the National Science Foundation through grant SES-1947662.}}

\author{
    Matias D. Cattaneo\footnote{Department of Operations Research and Financial Engineering, Princeton University.}\and
    Gregory Fletcher Cox\footnote{Department of Economics, National University of Singapore.}\and
    Michael Jansson\footnote{Department of Economics, University of California at Berkeley.}\and
    Kenichi Nagasawa\footnote{Department of Economics, University of Warwick.}
}
\maketitle

\setcounter{page}{0}\thispagestyle{empty}

\begin{abstract}
    Certain extremum estimators have asymptotic distributions that are non-Gaussian, yet characterizable as the distribution of the $\argmax$ of a Gaussian process. This paper presents high-level sufficient conditions under which such asymptotic distributions admit a continuous distribution function. The plausibility of the sufficient conditions is demonstrated by verifying them in three examples, namely maximum score estimation, empirical risk minimization, and threshold regression estimation. In turn, the continuity result buttresses several recently proposed inference procedures whose validity seems to require a result of the kind established herein. A notable feature of the high-level assumptions is that one of them is designed to enable us to employ the Cameron-Martin theorem. In a leading special case, the assumption in question is demonstrably weak and appears to be close to minimal.
\end{abstract}


\clearpage


\section{Introduction}

Certain extremum estimators have asymptotic distributions that are non-Gaussian, yet characterizable as the distribution of the $\argmax$ of a Gaussian process. To fix ideas, letting $\Btheta_0\in\R^d$ denote a parameter (vector) of interest, the estimators $\hat{\Btheta}_n$ in question satisfy
\begin{equation}\label{equation: Chernoff-type asymptotics}
    r_n(\hat{\Btheta}_n-\Btheta_0)\rightsquigarrow \argmax_{\Bs\in\R^d}\CG(\Bs),
\end{equation}
where $n$ is the sample size, $r_n$ is a rate of convergence, $\rightsquigarrow$ denotes weak convergence (as $n\to\infty$), and $\CG$ is a Gaussian process admitting a unique  maximizer (over $\R^d$) whose distribution is non-Gaussian. The seminal work of \cite{Kim-Pollard_1990_AoS} was concerned with (cube root asymptotic) cases where $r_n=\sqrt[3]{n}$ and the mean function of $\CG$ is a quadratic form, but subsequent work \citep*[e.g.,][]{Hansen_2000_ECMA,Lai-Lee_2005_JASA,Lee-Liao-Seo-Shin_2021_AoS,Lee-Pun_2006_JASA,Lee-Yang_2020_AoS,Westling-Carone_2020_AoS,Yu-Fan_2021_JBES} has documented the relevance of allowing for the extra flexibility afforded by the more general formulation in \eqref{equation: Chernoff-type asymptotics}.

Letting $\mu$ and $\CC$ denote the mean function and covariance kernel of $\CG$ and defining
\begin{equation}\label{equation: Chernoff-type cdf}
    F_{\hat{\Bs}}(\Bt)=\P[\hat{\Bs}\leq \Bt], \qquad \hat{\Bs}=\argmax_{\Bs\in\R^d}\CG(\Bs),
\end{equation}
our goal in this paper is to give conditions on $\mu$ and $\CC$ that imply continuity of $F_{\hat{\Bs}}$. Continuity of $F_{\hat{\Bs}}$ is useful when the goal is to use $\hat{\Btheta}_n$ to construct confidence regions. For instance, \citet[Lemma 23.3]{vanderVaart_1998_book} assumes continuity when establishing validity of bootstrap-based confidence intervals; see also \citet*[Section 1.2]{Politis-Romano-Wolf_1999_Book}. Moreover, and relatedly, it follows from Polya's theorem that if \eqref{equation: Chernoff-type asymptotics} holds and if $F_{\hat{\Bs}}$ is continuous, then the the probability laws of $r_n(\hat{\Btheta}_n-\Btheta_0)$ converge to the law with distribution function $F_{\hat{\Bs}}$ not only in the bounded Lipschitz metric (or any other metric metrizing weak convergence), but also in the Kolmogorov metric; that is, we have a result of the form
\begin{equation}\label{equation: Chernoff-type asymptotics, Kolmogorov}
    \sup_{\Bt\in\R^d}\left|\P\Big[r_n(\hat{\Btheta}_n-\Btheta_0)\leq\Bt\Big]-F_{\hat{\Bs}}(\Bt)\right|\to0.
\end{equation}

When $\mu$ is a quadratic form and $\CC$ is a bilinear form, the distribution of $\hat{\Bs}$ is Gaussian. More generally, under mild conditions on $\mu$ the distribution of $\hat{\Bs}$ is that of a transformation of a Gaussian vector when $\CC$ is a bilinear form, implying in particular that the properties of $F_{\hat{\Bs}}$ can be deduced by means of a change of variables argument. Two other special cases where a complete characterization of $F_{\hat{\Bs}}$ is available are when $d=1$, $\CC$ is the covariance kernel of a two-sided Brownian motion, and $\mu$ is proportional to either the absolute value function or the square function. In both cases, the distribution of $\hat{\Bs}$ is that of a scalar multiple of a random variable with a well-known continuous distribution. Somewhat more generally, \citet*[Lemma A.2]{Cattaneo-Jansson-Nagasawa_2024_AoS} gave conditions on $\mu$ under which $F_{\hat{\Bs}}$ is continuous when $d=1$ and $\CC$ is the covariance kernel of a two-sided Brownian motion. On the other hand, little (if anything) appears to be known about the properties of $F_{\hat{\Bs}}$ when $d>1$ and $\CC$ is not bilinear.

In this paper we close this gap by presenting sufficient conditions for continuity of $F_{\hat{\Bs}}$ that do not require $d=1$ and are applicable (only) when $\CC$ is not bilinear. Proceeding under the assumption that $d=1$, the proof of \citet[Lemma A.2]{Cattaneo-Jansson-Nagasawa_2024_AoS} establishes continuity of $F_{\hat{\Bs}}$ by showing that the distribution of $\hat{\Bs}$ is atomless (under the additional assumptions of the lemma). The method of proof can be adapted to give conditions under which the distribution of $\hat{\Bs}$ is atomless also when $d>1$, but when $d>1$ a distribution can be atomless even if the associated distribution function is discontinuous. Establishing continuity of $F_{\hat{\Bs}}$ when $d>1$ therefore requires a fundamentally different method of proof than that employed by \citet[Lemma A.2]{Cattaneo-Jansson-Nagasawa_2024_AoS}. The differences in proof strategies are reflected also in the assumptions under which the proofs proceed. Notably, one of the conditions imposed in this paper explicitly involves both $\mu$ and $\CC$ and requires that for every $N\in\N$, restriction of $\CG$ to $[-N,N]^d$ has a mean function that belongs to the reproducing kernel Hilbert space (RKHS) of its covariance kernel. By the Cameron-Martin theorem, if a Gaussian process has a mean belonging to the RKHS of its covariance kernel, then its induced probability measure and the probability measure induced by its centered version are mutually absolutely continuous. The proof of our main result uses this fact and an assumed shift equivariance property of the covariance kernel to deduce continuity of $F_{\hat{\Bs}}$.

The usefulness of our main result is illustrated by applying it to three examples: maximum score estimation, empirical risk minimization, and threshold regression estimation. Each example involves an estimator satisfying \eqref{equation: Chernoff-type asymptotics} with $d$ possibly greater than one and a covariance kernel that is not bilinear. Although distinct in several ways, the examples enjoy the common feature that continuity of $F_{\hat{\Bs}}$ can be shown by verifying the conditions of our main result. In particular, the condition that the mean function belongs to the RKHS of the covariance kernel can be verified by following a general strategy outlined in \cite{Lifshits_1995_Book}. 

In addition to facilitating the justification of certain large-sample inference procedures based on distributional approximations of the form \eqref{equation: Chernoff-type asymptotics}, our paper sheds new light on the canonical problem of characterizing the distributional properties of the $\argmax$ of a Gaussian process. That problem is substantially different from the well-studied problem of understanding the distributional properties of the maximum itself, where the $d=1$ case is mostly settled \citep[e.g.,][and references therein]{Lifshits_1995_Book}, the multidimensional case is fairly well understood \citep[e.g.,][and references therein]{Azais-Wschebor_2005_AOAP}, and where, more generally, continuity of the distribution function of the maximum can be established with the help of anti-concentration results \citep*[e.g.,][]{Chernozhukov-Chetverikov-Kato_2015_PTRF}. However, as noted by \citet[p. 3494]{Samorodnitsky-Shen_2013_AOP}, ``very little is known about the random location of the supremum'' of a Gaussian process. In the multidimensional case, we are only aware of \cite{Azais-Chassan_2020_SPA}, which shows that the distribution admits a density under the assumption that the sample paths are twice differentiable.

The remainder of the paper proceeds as follows. Section \ref{Section: Motivating Examples} introduces our three examples. Our main result is presented in Section \ref{Section: Main Result}, while Section \ref{Section: Verification of Assumption 1} outlines a general strategy for verifying the conditions of the main result and demonstrates how to apply it in the examples. Finally, Section \ref{Section: Discussion of Assumption 1(iii)} compares our results with the known results alluded to in the third paragraph of this section.

\section{Motivating Examples}\label{Section: Motivating Examples}

The class of estimators satisfying \eqref{equation: Chernoff-type asymptotics} is rich, containing examples in econometrics, statistics, and other data science disciplines. To further motivate our work, this section presents three representative examples.

\subsection{Maximum Score}\label{Section: Maximum Score}

Suppose $\{(y_i,w_i,\Bx_i')'\}_{i=1}^n$ is a random sample from the distribution of a vector $(y,w,\Bx')'$ generated by the semiparametric binary response model
\begin{equation*}
    y = \I\{w + \Bx'\Btheta_0 \geq u \},\qquad \text{Median}(u|w,\Bx)=0,
\end{equation*}
where $\I\{\cdot\}$ is the indicator function, $w,u\in\R$ and $\Bx\in\R^d$ are random variables, and $\Btheta_0\in\Theta\subseteq\R^d$ is the parameter of interest. \cite{Manski_1975_JoE} introduced the maximum score estimator of $\Btheta_0$, which is any maximizer $\hat\Btheta_n$ of
\begin{equation*}
    \sum_{i=1}^n (2y_i-1)\I\{w_i+\Bx_i'\Btheta \geq 0\}
\end{equation*}
with respect to $\Btheta\in\Theta$. Using the methods of \cite{Kim-Pollard_1990_AoS}, \cite{Abrevaya-Huang_2005_ECMA} gave regularity conditions under which \eqref{equation: Chernoff-type asymptotics} holds with $r_n=\sqrt[3]{n}$ and $\CG$ being a Gaussian process whose mean function and covariance kernel take the form
\begin{equation*}
    \mu(\Bs)=-\Bs'\E\left[f_{u|w,\Bx}(0|-\Bx'\Btheta_0,\Bx)f_{w|\Bx}(-\Bx'\Btheta_0|\Bx)\Bx\Bx'\right]\Bs
\end{equation*}
and
\begin{equation*}
    \CC(\Bs,\Bt)=\E\left[f_{w|\Bx}(-\Bx'\Btheta_0|\Bx)\CC_{\mathtt{BM}}(\Bx'\Bs,\Bx'\Bt)\right],
\end{equation*}
respectively, where $f_{u|w,\Bx}$ and $f_{w|\Bx}$ denote conditional (Lebesgue) densities, and where $\CC_{\mathtt{BM}}$ is the covariance kernel of a two-sided standard Brownian motion; that is,
\begin{equation*}
    \CC_{\mathtt{BM}}(s,t)=\min\{|s|,|t|\}\I\{\sgn(s)=\sgn(t)\},
\end{equation*}
with $\sgn(\cdot)$ denoting the sign function.

When $d=1$, because $\mu$ is quadratic and $\CC$ is a scalar multiple of $\CC_{\mathtt{BM}}$, it follows from  \citet[Exercise 3.2.5]{vanderVaart-Wellner_2023_book} that the distribution of $\hat{\Bs}$ is that of a scalar multiple of a random variable with a well-known continuous distribution, namely the \citet{Chernoff_1964_AISM} distribution.
For $d>1$, on the other hand, it would appear to be an open question whether $F_{\hat{\Bs}}$ is continuous. We provide an affirmative answer to that question below, hereby buttressing a variety of inference procedures based on the maximum score estimator.

For specificity, consider the procedure of \cite*{Cattaneo-Jansson-Nagasawa_2020_ECMA}. That paper proposed a bootstrap-based estimator $\tilde{\Btheta}_n^*$ and gave conditions under which this estimator satisfies
\begin{equation*}
    r_n(\tilde{\Btheta}_n^*-\hat{\Btheta}_n)\rightsquigarrow_\P \argmax_{\Bs\in\R^d}\CG(\Bs),
\end{equation*}
where $\rightsquigarrow_\P$ denotes weak convergence in probability. Because $F_{\hat{\Bs}}$ is continuous, the displayed result can be combined with \eqref{equation: Chernoff-type asymptotics} to yield a bootstrap consistency result of the form
\begin{equation*}
    \sup_{\Bt\in\R^d}\left|\P_n^*\left[\tilde{\Btheta}_n^*-\hat{\Btheta}_n\leq\Bt\right]- \P\left[\hat{\Btheta}_n-\Btheta_0\leq\Bt\right]\right|\to_\P0,
\end{equation*}
where $\P_n^*$ is the bootstrap probability measure. As a consequence, for any $\Blambda\in\R^d$, defining
\begin{equation*}
    q_{\Blambda,n}^*(t) = \inf\left\{ q \in \R : \P_n^* [\Blambda'\tilde{\Btheta}_n^* - \Blambda'\hat{\Btheta}_n \leq q] \geq t \right\}, \qquad t\in(0,1),
\end{equation*}
\citet[Lemma 23.3]{vanderVaart_1998_book} shows that the equal-tailed ``percentile'' interval
\begin{equation*}
    \mathsf{CI}_{\Blambda,n}^*(1-\alpha) = \left[\Blambda'\hat{\Btheta}_n - q_{\Blambda,n}^*(1-\alpha/2) ~,~ \Blambda'\hat{\Btheta}_n - q_{\Blambda,n}^*(\alpha/2) \right]
\end{equation*}
is a confidence interval (for $\Blambda'\Btheta_0$) of asymptotic level $1-\alpha$:
\begin{equation*}
    \lim_{n \to \infty} \P\left[\Blambda'\Btheta_0 \in \mathsf{CI}_{\Blambda,n}^*(1 - \alpha)\right] = 1 - \alpha.
\end{equation*}
With minor modifications, similar comments apply to the inference procedures proposed by \cite*{Delgado-RodriguezPoo-Wolf_2001_EL}, \cite{Hong-Li_2020_AoS}, \cite*{Jun-Pinkse-Wan_2015_JoE}, \cite{Lee-Yang_2020_AoS}, and \cite*{Patra-Seijo-Sen_2018_JoE}.

Collectively, the inference procedures mentioned in the previous paragraph therefore constitute asymptotically valid alternatives to inference procedures based on the smoothed maximum score estimator of \cite{Horowitz_1992_ECMA}. (A finite-sample inference procedure for the semiparametric binary response model has recently been proposed by \cite{Rosen-Ura_2025_REStud}.)

\subsection{Empirical Risk Minimization}\label{Section: Empirical Risk Minimization}

\cite{Mohammadi-vandeGeer_2005_JMLR} considered the classification problem of estimating the minimizer $\Btheta_0\in\Theta\subseteq\R^d$ of the classification error $\P[y\neq h_{\Btheta}(x)]$ with respect to $\Btheta\in\Theta$, where $y\in \{-1,1\}$ is a binary outcome, $x\in\CX\subseteq\R$ is a scalar feature, and $\{h_{\Btheta}:\Btheta\in\Theta\}$ is a collection of classifiers. Given a random sample $\{(y_i,x_i)\}_{i=1}^n$ from the distribution of $(y,x)$, an empirical risk minimizer is a minimizer $\hat{\Btheta}_n$  of
\begin{equation*}
    \sum_{i=1}^n \I\{y_i\neq h_{\Btheta}(x_i) \}.
\end{equation*}

Setting $\CX=[0,1]$ and specializing to the case where the classifiers are of the form
\begin{equation*}
    h_{\Btheta}(x) = \sum_{\ell=1}^{d+1} (-1)^\ell \I\{\theta_{\ell-1}\leq x<\theta_{\ell}\}
\end{equation*}
for $\Btheta=(\theta_1,\dots,\theta_d)'\in\Theta=\{\Btheta\in [0,1]^d: 0=\theta_0\leq \theta_1\leq \dots\leq \theta_d\leq \theta_{d+1}=1\}$, \citet[Theorem 1]{Mohammadi-vandeGeer_2005_JMLR} gave conditions under which \eqref{equation: Chernoff-type asymptotics} holds with $r_n=\sqrt[3]{n}$ and $\CG$ being a Gaussian process whose mean function and covariance kernel take the form
\begin{equation}\label{equation: ERM mean function}
    \mu(\Bs)= \sum_{\ell=1}^d \mu_\ell(s_\ell), \qquad \mu_\ell(s_\ell)=(-1)^{\ell}p(\theta_{0,\ell})f(\theta_{0,\ell}) s_\ell^2,
\end{equation}
and
\begin{equation}\label{equation: ERM covariance kernel}
    \CC(\Bs,\Bt)=\sum_{\ell=1}^d \CC_\ell(s_\ell,t_\ell), \qquad \CC_\ell(s_\ell,t_\ell)=f(\theta_{0,\ell})\CC_{\mathtt{BM}}(s_\ell,t_\ell),
\end{equation}
respectively, where $\Btheta_0=(\theta_{0,1},\dots,\theta_{0,d})',\Bs=(s_1,\dots,s_d)',\Bt=(t_1,\dots,t_d)'$, $f$ is a Lebesgue density of $x$, $p(x)= \text{d} \P[y=1|x]/\text{d}x$, and where the assumptions imposed on the model ensure that $(-1)^{\ell}p(\theta_{0,\ell})f(\theta_{0,\ell})<0$ for every $\ell=1,\dots,d$.

This example is similar to the maximum score example insofar as when $d=1$, the distribution of $\hat{\Bs}$ is that of a scalar multiple of a random variable with a Chernoff distribution. In fact, also when $d>1$, the elements of $\hat{\Bs}=(\hat{s}_1,\dots,\hat{s}_d)'$ are mutually independent, each having a distribution which is that of a scalar multiple of a random variable with a Chernoff distribution. Indeed, letting $\CG_1,\dots,\CG_d$ be mutually independent Gaussian processes with mean functions $\mu_1,\dots,\mu_d$ and covariance kernels $\CC_1,\dots,\CC_d$, respectively, $\CG$ admits the representation $\CG(\Bs)=\sum_{\ell=1}^d \CG_\ell(s_\ell)$, implying in particular that
\begin{equation*}
    \hat{s}_\ell=\argmax_{s_\ell\in\R}\CG_\ell(s_\ell)\qquad \text{for each }\ell\in\{1,\dots,d\}.
\end{equation*}

In other words, this example has special structure that can be used to obtain a definitive characterization of the distribution of $\hat{\Bs}$ without utilizing new tools. It is nevertheless of interest to explore the ease with which the technology developed in this paper can be deployed to establish continuity of $F_{\hat{\Bs}}$ in this example. In particular, it is of interest to explore whether the (effective) ``dimension reduction'' permitted by this example can be leveraged when verifying the conditions of Theorem \ref{Theorem} below.

\subsection{Threshold Regression}\label{Section: Threshold Regression}

Consider the threshold regression model 
\begin{equation*}
    y=\Bx'\Bbeta_0+\Bx'\Bdelta_n\I\{q>\Bw'\Btheta_0\}+u,\qquad \mathbb{E}(u|\Bx, q, \Bw)=0,  
\end{equation*}
where $y\in\R$ is a dependent variable, $\Bx\in\R^k$ is a (possibly) vector-valued regressor, $q$ is a threshold variable, $\Bw\in\R^d$ is a (possibly) vector-valued factor governing the threshold cutoff, and where, borrowing ideas from the change-point literature \citep[e.g.,][]{Bai_1997_REStat}, $\Bdelta_n$ is a ``threshold effect'' whose magnitude vanishes with $n$. The present model (as well as distinct generalizations thereof) has been studied by \cite*{Lee-Liao-Seo-Shin_2021_AoS} and \cite{Yu-Fan_2021_JBES}, and differs from the model considered in \cite{Hansen_2000_ECMA} by allowing the factor $\Bw$ to be non-constant.

Given a random sample $\{(y_i,\Bx_i',q_i,\Bw_i')'\}_{i=1}^n$ from the distribution of $(y,\Bx',q,\Bw')'$, a least squares estimator $(\hat{\Bbeta}_n',\hat{\Bdelta}_n',\hat{\Btheta}_n')'$ of $(\Bbeta_0',\Bdelta_n',\Btheta_0')'$ is a minimizer of
\begin{equation*}
    \sum_{i=1}^n \left(y_i-\Bx_i'\Bbeta-\Bx_i'\Bdelta\I\{q_i>\Bw_i'\Btheta\}\right)^2
\end{equation*}
over $(\Bbeta',\Bdelta',\Btheta')'\in\R^{2k+d}$. Assuming $\|\Bdelta_n\|\rightarrow 0$, $n\|\Bdelta_n\|^2\rightarrow\infty$, and $\bar{\Bdelta}_n=\Bdelta_n/\|\Bdelta_n\|\rightarrow \bar{\Bdelta}$ (for some $\bar{\Bdelta}\in \S^{d-1}=\{\bar{\Bdelta}\in\R^d:\|\bar{\Bdelta}\|=1\}$), \cite{Yu-Fan_2021_JBES} gave conditions under which \eqref{equation: Chernoff-type asymptotics} holds with $r_n=n\|\Bdelta_n\|^2$ and $\CG=\CG(\cdot;\bar{\Bdelta})$ being a Gaussian process whose mean function and covariance kernel take the form
\begin{equation*}
    \mu(\Bs)=\mu(\Bs;\bar{\Bdelta})=-\frac{1}{2}\bar{\Bdelta}'\E\left[|\Bw'\Bs|f_{q|\Bw}(\Bw'\Btheta_0|\Bw)\E_{\cdot|q,\Bw}[\Bx\Bx'|\Bw'\Btheta_0,\Bw]\right]\bar{\Bdelta}
\end{equation*}
and
\begin{equation*}
    \CC(\Bs,\Bt)=\CC(\Bs,\Bt;\bar{\Bdelta})=\bar{\Bdelta}'\E\left[f_{q|\Bw}(\Bw'\Btheta_0|\Bw)\E_{\cdot|q,\Bw}[\Bx\Bx'u^2|\Bw'\Btheta_0,\Bw]\CC_{\mathtt{BM}}(\Bw'\Bs,\Bw'\Bt)\right]\bar{\Bdelta}, 
\end{equation*}
respectively, where $\E_{\cdot|q,\Bw}$ denotes conditional expectation.

When $d=1$, the distribution of $\hat{\Bs}$ is that of a scalar multiple of a random variable with a (known) continuous distribution \citep[e.g., Section 3.2 of][]{Hansen_2000_ECMA}. For $d>1$, on the other hand, it would appear to be an open question whether $F_{\hat{\Bs}}$ is continuous. We provide an affirmative answer to that question below.

\paragraph*{Remark.} 
As a by-product, continuity of the limiting distribution function can be used to show that if $\|\Bdelta_n\|\rightarrow 0$ and if $n\|\Bdelta_n\|^2\rightarrow\infty$, then (whether or not $\bar{\Bdelta}_n$ is convergent in $\S^{d-1}$) we have
\begin{equation*}
    \sup_{t\in\R^d}\left|\P\left[r_n(\hat{\Btheta}_n-\Btheta_0)\leq\Bt\right]-\P\left[\argmax_{\Bs\in\R^d}\CG(\Bs;\bar{\Bdelta}_n)\leq\Bt\right]\right|\to0.
\end{equation*}

\section{Main Result}\label{Section: Main Result}

As before, let $\CG$ be a Gaussian process on $\R^d$ with mean function $\mu$ and covariance kernel $\CC$. Also, for any $N\in\N$, let $\CG_N$ be the restriction of $\CG$ to $[-N, N]^d$, let $\mu_N$ and $\CC_N$ be the mean function and covariance kernel of $\CG_N$, and let $\mathscr{H}_N$ be the RKHS of $\CC_N$ \cite[as defined in Section 2.6.1 of][]{Gine-Nickl_2016_Book}.

The following high-level assumption holds in each of the examples of Section \ref{Section: Motivating Examples}.

\begin{assumption}\label{Assumption}
    \hspace{1pt}
    \begin{enumerate}[label=(\roman*)]
        \item \label{Assumption: Existence of argmax}
        With probability one, $\CG$ has continuous sample paths and admits a maximizer over $\R^d$.
        
        \item \label{Assumption: Shift equivariance}
        For any $\Bs,\Bt\in\R^d$ and any $\Bh\in\R^{d}\backslash\{\mathbf{0}\}$, $\CC(\Bh,\Bh)>0$ and
        \begin{equation*}
            \CC(\Bh+\Bs,\Bh+\Bt)-\CC(\Bh+\Bs,\Bh)-\CC(\Bh,\Bh+\Bt)+\CC(\Bh,\Bh)=\CC(\Bs,\Bt).
        \end{equation*}
        
        \item \label{Assumption: RKHS}
        For any $N\in\N$, $\mu_N\in \mathscr{H}_N$. 
    \end{enumerate}
\end{assumption}

Part \textit{\ref{Assumption: Existence of argmax}} guarantees existence (with probability one) of a maximizer of $\CG$ over $\R^d$, and over any compact set $S\subset\R^d$. By \citet[Lemma 2.6]{Kim-Pollard_1990_AoS}, these maximizers are unique provided $\V[\CG(\Bs)-\CG(\Bt)]\neq0$ for every $\Bs\neq\Bt$. Part \textit{\ref{Assumption: Shift equivariance}} gives sufficient conditions for this non-degeneracy condition to hold and furthermore implies that the centered process $\CG^\mu=\CG-\mu$ is shift equivariant in the sense that the law of the process $\CG^\mu(\Bh+\cdot)-\CG^\mu(\Bh)$ is the same for every $\Bh\in\R^d$. An immediate implication of shift equivariance is that $\V[\CG(0)]=0$. A slightly more subtle implication is recorded in the following lemma.

\begin{lemma}\label{Lemma}
    Suppose $\CG$ has continuous sample paths and suppose Assumption \ref{Assumption}\textit{\ref{Assumption: Shift equivariance}} holds. For any $\Bh\in\mathbb{R}^d$, any measurable set $T\subseteq\mathbb{R}^d$, and any compact set $S\subset\mathbb{R}^d$,
    \begin{equation*}
        \P\left[\argmax_{\Bs\in S} \CG^\mu(\Bs) \in T\right] = \P\left[\argmax_{\Bs\in S+\Bh} \CG^\mu(\Bs) \in T +\Bh\right].
    \end{equation*}
\end{lemma}
\begin{myproof}{Lemma \ref{Lemma}}
    By change of variables and shifting by a constant,
    \begin{equation*}
        \argmax_{\Bs\in S +\Bh} \CG^\mu(\Bs)
        =\argmax_{\Bs\in S} \CG^\mu(\Bh+\Bs) +\Bh 
        =\argmax_{\Bs\in S} \{\CG^\mu(\Bh+\Bs)-\CG^\mu(\Bh)\}+\Bh.
    \end{equation*}
    The desired conclusion follows from shift equivariance of $\CG^\mu$.
\end{myproof}

By the Cameron-Martin theorem \citep[e.g., Theorem 2.6.13 of][]{Gine-Nickl_2016_Book}, part \textit{\ref{Assumption: RKHS}} of Assumption \ref{Assumption} ensures that the probability measures associated with $\CG_N$ and $\CG^\mu_N=\CG_N-\mu_N$ are mutually absolutely continuous for any $N\in\N$. Along with Lemma \ref{Lemma}, that property plays a key role in our proof of the following result.

\begin{theorem}\label{Theorem}
    Under Assumption \ref{Assumption}, $F_{\hat{\Bs}}$ in \eqref{equation: Chernoff-type cdf} is continuous.
\end{theorem}

\begin{myproof}{Theorem \ref{Theorem}}
    For any compact set $S\subset\mathbb{R}^d$, parts \textit{\ref{Assumption: Existence of argmax}} and \textit{\ref{Assumption: Shift equivariance}} of Assumption \ref{Assumption} guarantee existence (with probability one) of a unique maximizer of $\CG^\mu$ over $S$, uniqueness being a consequence of \citet[Lemma 2.6]{Kim-Pollard_1990_AoS}. This observation will be used repeatedly without further mention.
    The (joint) distribution function of $\hat{\Bs}=(\hat{s}_1,\dots,\hat{s}_d)'$ is continuous if and only if each of its marginal distribution functions is continuous. Fixing $\ell\in\{1,\dots,d\}$ and $t\in\R$, the proof can therefore be completed by showing that
    \begin{equation}\label{equation: continuity}
        \P\left[\hat{s}_\ell=t\right]=0.
    \end{equation}
    
    Defining $N_t=\lceil|t|\rceil+1$, letting $\Be_\ell$ denote the $\ell$th standard basis vector of $\R^d$, and noting that
    \begin{equation*}
        \left\{\hat{s}_\ell=t\right\} = \left\{\Be_\ell'\argmax_{\Bs\in\R^d}\CG(\Bs)=t\right\} \subseteq
        \bigcup_{N=N_t}^{\infty} \left\{\Be_\ell'\argmax_{\Bs\in[-N,N]^d}\CG(\Bs)=t\right\},
    \end{equation*}
    a sufficient condition for \eqref{equation: continuity} to hold is that
    \begin{equation*}
        \P\left[\Be_\ell'\argmax_{\Bs\in[-N,N]^d}\CG(\Bs)=t\right]=0 \qquad \text{for every } N \geq N_t.
    \end{equation*}
    By the Cameron-Martin theorem, under Assumption \ref{Assumption}\textit{\ref{Assumption: RKHS}} the displayed condition is equivalent to
    \begin{equation}\label{equation: continuity*}
        \P\left[\Be_\ell'\argmax_{\Bs\in[-N,N]^d}\CG^\mu(\Bs)=t\right]=0 \qquad \text{for every } N \geq N_t.
    \end{equation}
    
    Fixing $N\geq N_t$ and $J\geq 2$, let
    \begin{equation*}
        S_j=\left\{(s_1,\dots,s_d)\in[-N,N]^d:-N_t+\frac{1}{2}\frac{j-1}{J-1} \leq s_\ell \leq N_t+\frac{1}{2}\left(\frac{j-1}{J-1}-1\right)\right\} 
    \end{equation*}
    for $j\in\{1,\dots,J\}$. Noting that
    \begin{align*}
        [-N,N]^d &\supseteq \bar{S} = \{(s_1,\dots,s_d)\in[-N,N]^d:-N_t \leq s_\ell \leq N_t\}=\cup_{j=1}^{J}S_j\\
        &\supset\cap_{j=1}^{J}S_j=\{(s_1,\dots,s_d)\in[-N,N]^d:-N_t+1/2 \leq s_\ell \leq N_t-1/2\}=\underline{S}, 
    \end{align*}
    we have
    \begin{align*}
        \P\left[\Be_\ell'\argmax_{\Bs\in[-N,N]^d}\CG^\mu(\Bs)=t\right]
        &\leq \P\left[\Be_\ell'\argmax_{\Bs\in\bar{S}}\CG^\mu(\Bs)=t\right]
        \leq \P\left[\Be_\ell'\argmax_{\Bs\in S_1}\CG^\mu(\Bs)=t\right]\\
        &= \frac{1}{J}\sum_{j=1}^{J}\P\left[\Be_\ell'\argmax_{\Bs\in S_j}\CG^\mu(\Bs)=t+\frac{1}{2}\frac{j-1}{J-1}\right]\\
        &\leq \frac{1}{J}\sum_{j=1}^{J}\P\left[\Be_\ell'\argmax_{\Bs\in \underline{S}}\CG^\mu(\Bs)=t+\frac{1}{2}\frac{j-1}{J-1}\right]\\
        &= \frac{1}{J}\P\left[\Be_\ell'\argmax_{\Bs\in \underline{S}}\CG^\mu(\Bs)\in\left\{t+\frac{1}{2}\frac{j-1}{J-1}:1\leq j \leq J\right\}\right]\leq \frac{1}{J},
    \end{align*}
    where the first equality uses Lemma \ref{Lemma} and the second equality uses uniqueness of the maximizer of $\CG^\mu$ over $\underline{S}$. Since $J \geq 2$ was arbitrary, \eqref{equation: continuity*} follows.
\end{myproof}


\section{Verification of Assumption \ref{Assumption}}\label{Section: Verification of Assumption 1}

\subsection{Assumption \ref{Assumption}\textit{\ref{Assumption: Existence of argmax}}}

The continuity part of Assumption \ref{Assumption}\textit{\ref{Assumption: Existence of argmax}} is mild and usually trivial to verify. Under continuity and assuming that $\P[\CG(\mathbf{0})=0]=1$, a high-level sufficient condition for existence of a maximizer of $\CG$ over $\R^d$ is that
\begin{equation}\label{equation: argmax existence}
    \P\left[\limsup_{\|\Bs\|\rightarrow\infty}\CG(\Bs)<0\right]=1.
\end{equation}
In turn, proceeding as in the proof of \citet[Lemma 2.5]{Kim-Pollard_1990_AoS} it can be shown that if the covariance kernel satisfies the (self-similarity) property that for some $H>0$,
\begin{equation}\label{equation: argmax existence, covariance}
    \CC(\tau\Bs,\tau\Bt)=\tau^{2H}\CC(\Bs,\Bt) \qquad \text{for every }\Bs,\Bt\in\R^d,\tau>0,
\end{equation}
then \eqref{equation: argmax existence} is implied by the following mild condition on the mean function:
\begin{equation}\label{equation: argmax existence, mean}
    \limsup_{\|\Bs\|\rightarrow\infty}\frac{\mu(\Bs)}{\|\Bs\|^{H+\epsilon}}<0 \qquad \text{for some }\epsilon>0.
\end{equation}
The assumption $\P[\CG(\mathbf{0})=0]=1$ holds if (and only if) $\mu(\mathbf{0})=0=\CC(\mathbf{0},\mathbf{0})$ and is therefore satisfied in each of the examples of Section \ref{Section: Motivating Examples}. Likewise, the conditions \eqref{equation: argmax existence, covariance} and \eqref{equation: argmax existence, mean} are both fairly primitive. Also, by inspection, \eqref{equation: argmax existence, covariance} can be seen to hold with $H=1/2$ in each of the examples of Section \ref{Section: Motivating Examples}. Moreover, setting $H=1/2$, \eqref{equation: argmax existence, mean} can be seen to hold with $\epsilon=3/2$ in the maximum score and empirical risk minimization examples, and with $\epsilon=1/2$ in the threshold regression example.

To explain why it is no coincidence that the self-similarity property of $\CC$ holds in the examples of Section \ref{Section: Motivating Examples}, it may be helpful to note that in each case $\CG$ is the weak limit of a process of the form
\begin{equation*}
    \Bs\mapsto\sqrt{\frac{r_n}{n}}\sum_{i=1}^{n}[m_n(\Bz_i,\Btheta_0+r_n^{-1}\Bs)-m_n(\Bz_i,\Btheta_0)],
\end{equation*}
where $\{\Bz_i\}_{i=1}^n$ is a random sample and $m_n$ is some function (possibly depending on $n$). For instance, in the threshold regression example we have
\begin{equation*}
    m_n(\Bz,\Btheta)=-\frac{1}{2\|\Bdelta_n\|}\left(y-\Bx'\Bbeta_0-\Bx'\Bdelta_n\I\{q>\Bw'\Btheta\}\right)^2,\qquad \Bz=(y,\Bx',q,\Bw')'.
\end{equation*}
It therefore stands to reason that $\CC$ can be characterized as follows:
\begin{equation*}
    \CC(\Bs,\Bt)=\lim_{n\to\infty}r_n\E[\{m_n(\Bz,\Btheta_0+r_n^{-1}\Bs)-m_n(\Bz,\Btheta_0)\}\{m_n(\Bz,\Btheta_0+r_n^{-1}\Bt)-m_n(\Bz,\Btheta_0)\}].
\end{equation*}
To further conclude that \eqref{equation: argmax existence, covariance} holds with $H=1/2$, it suffices to assume that the preceding display admits the following strengthening: for any $\eta_n>0$ with $\eta_n=O(r_n^{-1})$, 
\begin{equation}\label{equation: pointwise convergence to covariance kernel}
    \CC(\Bs,\Bt) = \lim_{n\to\infty} \frac{\E[\{m_n(\Bz,\Btheta_0+\eta_n\Bs)-m_n(\Bz,\Btheta_0)\}\{m_n(\Bz,\Btheta_0+\eta_n\Bt)-m_n(\Bz,\Btheta_0)\}]}{\eta_n}.
\end{equation}
A characterization of the form \eqref{equation: pointwise convergence to covariance kernel} is valid in each of the examples of Section \ref{Section: Motivating Examples}.

\subsection{Assumption \ref{Assumption}\textit{\ref{Assumption: Shift equivariance}}}

By inspection, the displayed part of Assumption \ref{Assumption}\textit{\ref{Assumption: Shift equivariance}} holds in each of the examples of Section \ref{Section: Motivating Examples}. To explain why this is no coincidence, observe that upon defining $\Btheta_n=\Btheta_0+\eta_n\Bh$ we have
\begin{align*}
    &\E[\{m_n(\Bz,\Btheta_0+\eta_n[\Bh+\Bs])-m_n(\Bz,\Btheta_0)\}\{m_n(\Bz,\Btheta_0+\eta_n[\Bh+\Bt])-m_n(\Bz,\Btheta_0)\}] \\
    &\quad - \E[\{m_n(\Bz,\Btheta_0+\eta_n[\Bh+\Bs])-m_n(\Bz,\Btheta_0)\}\{m_n(\Bz,\Btheta_0+\eta_n\Bh)-m_n(\Bz,\Btheta_0)\}] \\
    &\quad - \E[\{m_n(\Bz,\Btheta_0+\eta_n\Bh)-m_n(\Bz,\Btheta_0)\}\{m_n(\Bz,\Btheta_0+\eta_n[\Bh+\Bt])-m_n(\Bz,\Btheta_0)\}] \\
    &\quad + \E[\{m_n(\Bz,\Btheta_0+\eta_n\Bh)-m_n(\Bz,\Btheta_0)\}\{m_n(\Bz,\Btheta_0+\eta_n\Bh)-m_n(\Bz,\Btheta_0)\}] \\
    &= \E[\{m_n(\Bz,\Btheta_n+\eta_n\Bs)-m_n(\Bz,\Btheta_n)\}\{m_n(\Bz,\Btheta_n+\eta_n\Bt)-m_n(\Bz,\Btheta_n)\}] \qquad \text{for any }\Bs,\Bt,\Bh\in\mathbb{R}^d.
\end{align*}
The displayed part of Assumption \ref{Assumption}\textit{\ref{Assumption: Shift equivariance}} is therefore valid whenever the following ``local uniform'' version of \eqref{equation: pointwise convergence to covariance kernel} is valid: for any $\eta_n>0$ with $\eta_n=O(r_n^{-1})$ and any $\Btheta_n=\Btheta_0+O(\eta_n)$,
\begin{equation}\label{equation: local uniform convergence to covariance kernel}
    \CC(\Bs,\Bt) = \lim_{n\to\infty} \frac{\E[\{m_n(\Bz,\Btheta_n+\eta_n\Bs)-m_n(\Bz,\Btheta_n)\}\{m_n(\Bz,\Btheta_n+\eta_n\Bt)-m_n(\Bz,\Btheta_n)\}]}{\eta_n}.
\end{equation}
In turn, a characterization of the form \eqref{equation: local uniform convergence to covariance kernel} is valid in each of the examples of Section \ref{Section: Motivating Examples}.

\subsection{Assumption \ref{Assumption}\textit{\ref{Assumption: RKHS}}}\label{Subsection: Assumption 1(iii)}

Evaluating Assumption \ref{Assumption}\textit{\ref{Assumption: RKHS}} is usually straightforward when $\mathscr{H}_N$ is known. More generally, a viable strategy for verifying Assumption \ref{Assumption}\textit{\ref{Assumption: RKHS}} can be based on \citet[Chapters 6 and 9]{Lifshits_1995_Book}. This subsection first outlines that strategy, and then demonstrates its usefulness by employing it in each of the examples in Section \ref{Section: Motivating Examples}.

\subsubsection{General Strategy}\label{General Strategy}

For $N\in\N$, suppose $\mathscr{E}_N=\{e_N(\cdot;\Bs):\Bs\in [-N,N]^d\}$ is a model of the covariance kernel $\CC_N$ \citep[in the terminology of Chapter 6 of][]{Lifshits_1995_Book}; that is, suppose that for some measure space $(\Omega_N,\CB_N,\nu_N)$, $\mathscr{E}_N$ is a collection of elements of $L_2(\Omega_N,\CB_N,\nu_N)$ satisfying
\begin{equation}\label{equation: RKHS representation covariance}
    \CC_N(\Bs,\Bt)=\int e_N(\Bomega;\Bs) e_N(\Bomega;\Bt)\text{d}\nu_N(\Bomega) \qquad \text{ for all } \Bs,\Bt\in[-N,N]^d.
\end{equation} 
Then, as discussed in \citet[Chapter 9]{Lifshits_1995_Book}, the mean function $\mu_N$ belongs to $\mathscr{H}_N$ if (and only if) it admits a function $l_N \in L_2(\Omega_N,\CB_N,\nu_N)$ satisfying
\begin{equation}\label{equation: RKHS representation mean}
    \mu_N(\Bs)=\int e_N(\Bomega;\Bs) l_N(\Bomega) \text{d}\nu_N(\Bomega) \qquad \text{ for all }\Bs\in[-N,N]^d.
\end{equation} 
\citet{Lifshits_1995_Book} demonstrates how this strategy can be used to characterize $\mathscr{H}_N$ for several examples of Gaussian processes. 
There is no general blueprint for defining $\mathscr{E}_N$ and $l_N$ satisfying \eqref{equation: RKHS representation covariance}-\eqref{equation: RKHS representation mean}. 
We modify the arguments used in \citet{Lifshits_1995_Book} to cover our examples. 

\subsubsection{Example: Maximum Score (Continued)}

For $N\in\N$, let $\CB_N$ be the Borel $\sigma$-algebra on $\Omega_N=\R^{1+d}$ and let $\nu_N = \lambda\times \P_\Bx$, where $\lambda$ is the Lebesgue measure on $\R$ and $\P_\Bx$ is the probability measure induced by $\Bx$. A direct calculation shows that \eqref{equation: RKHS representation covariance}-\eqref{equation: RKHS representation mean} hold with $\Bomega=(\omega_1,\Bx')'$,
\begin{equation*}
    e_N(\Bomega;\Bs) =\left[\I\{0 \leq \omega_1 \leq \Bx'\Bs\} + \I\{\Bx'\Bs\leq \omega_1 <0\} \right] \sqrt{f_{w|\Bx}(-\Bx'\Btheta_0|\Bx)},
\end{equation*}
and
\begin{equation*}
    l_N(\Bomega) = -2\I\{ |\omega_1|\leq N\sqrt{d}\|\Bx\| \}|\omega_1| f_{u|w,\Bx}(0|-\Bx'\Btheta_0,\Bx) \sqrt{f_{w|\Bx}(-\Bx'\Btheta_0|\Bx)}.
\end{equation*}

\subsubsection{Example: Empirical Risk Minimization  (Continued)}

For $N\in\N$, let $\CB_N$ be the Borel $\sigma$-algebra on $\Omega_N=\R^d$ and let $\nu_N$ be the Lebesgue measure on $\R^d$. Then \eqref{equation: RKHS representation covariance}-\eqref{equation: RKHS representation mean} hold with $\Bomega=(\omega_1,\dots,\omega_d)'$,
\begin{equation*}
    e_N(\Bomega;\Bs) = \sqrt{24Nu_N(\Bomega)}\sum_{\ell=1}^d \left(\omega_{\ell}-\frac{\sqrt[3]{s_{\ell}}}{2}\right)\left[ \I\{0\leq \omega_{\ell} \leq \sqrt[3]{s_{\ell}} \} + \I\{\sqrt[3]{s_{\ell}} \leq \omega_{\ell} <0 \} \right] \sqrt{f(\theta_{0,\ell})}
\end{equation*}
and
\begin{equation*}
    l_N(\Bomega) = \sqrt{\frac{75Nu_N(\Bomega)}{2}}\sum_{\ell=1}^d \omega_{\ell}^3|\omega_{\ell}|(-1)^{\ell} p(\theta_{0,\ell}) \sqrt{f(\theta_{0,\ell})},
\end{equation*}
where $u_N$ is a Lebesgue density of the uniform distribution on $[-N,N]^d$.

The strategy described in Section \ref{General Strategy} and followed in the previous paragraph is general and is not designed to leverage the special structure highlighted in Section \ref{Section: Empirical Risk Minimization}. Because it stands to reason that the additive separability of $\CC_N$ induces an analogous simplification of the associated $\mathscr{H}_N$, it seems natural to ask whether such a ``tensorization'' (is materialized and) can in turn be exploited when verifying Assumption \ref{Assumption}\textit{\ref{Assumption: RKHS}}.

The special structure \eqref{equation: ERM covariance kernel} of $\CC_N$ implies that $\mathscr{H}_N$ consists of those functions that are of the form $h_N(\Bs)= \sum_{\ell=1}^d h_{\ell,N}(s_\ell)$, with $h_{\ell,N}$ belonging to the RKHS of $\CC_{\ell,N}$ for each $\ell$. Now, proceeding as in \citet[Example 2.6.7]{Gine-Nickl_2016_Book} it can be shown that the RKHSs of $\CC_{1,N},\dots,\CC_{d,N}$ all (coincide and) consist of those functions on $[-N,N]$ that are zero-preserving and absolutely continuous with a square integrable (weak) derivative. In particular, $\mu_N\in\mathscr{H}_N$ when $\mu$ is given by \eqref{equation: ERM mean function}. 

\subsubsection{Example: Threshold Regression (Continued)}

For $N\in\N$, let $\CB_N$ be the Borel $\sigma$-algebra on $\Omega_N=\R^{1+d}$ and let $\nu_N = \lambda\times \P_{\Bw}$, where $\lambda$ is the Lebesgue measure on $\R$ and $\P_{\Bw}$ is the probability measure induced by $\Bw$. Then \eqref{equation: RKHS representation covariance}-\eqref{equation: RKHS representation mean} hold with $\Bomega=(\omega_1,\Bw')'$,
\begin{equation*}
    e_N(\Bomega;\Bs)=\left[\I\{\Bw'\Bs\leq \omega_1 <0\}+ \I\{0\leq\omega_1\leq\Bw'\Bs\}\right]\sqrt{\E_{\cdot|q,\Bw}[(\bar{\Bdelta}'\Bx u)^2|\Bw'\Btheta_0,\Bw]f_{q|\Bw}(\Bw'\Btheta_0|\Bw)}
\end{equation*}
and
\begin{align*}
    l_N(\Bomega)
    &= -\frac{1}{2}\I\{|\omega_1|\leq N\sqrt{d}\|\Bw\|\} \E_{\cdot|q,\Bw}[(\bar{\Bdelta}'\Bx)^2|\Bw'\Btheta_0,\Bw]\sqrt{\frac{f_{q|\Bw}(\Bw'\Btheta_0|\Bw)}{\E_{\cdot|q,\Bw}[(\bar{\Bdelta}'\Bx u)^2|\Bw'\Btheta_0,\Bw]}}.
\end{align*}

\section{Discussion of Assumption \ref{Assumption}\textit{\ref{Assumption: RKHS}}}\label{Section: Discussion of Assumption 1(iii)}

Interpreted as a condition on the mean function $\mu$, the strength of Assumption \ref{Assumption} \textit{\ref{Assumption: RKHS}} is inversely related to the richness of the RKHSs $\mathscr{H}_N$ generated by the covariance kernel $\CC$. It seems natural to ask, therefore, whether certain covariance kernels are so simple that Theorem \ref{Theorem} is silent about the continuity properties of $F_{\hat{\Bs}}$ for many (possibly most) interesting mean functions. Revisiting a particularly simple covariance kernel, Section \ref{Subsection: Bilinear Covariance Kernel} provides an affirmative answer to that question.

A related, but arguably more interesting, question is whether Assumption \ref{Assumption} \textit{\ref{Assumption: RKHS}} is likely to be ``close'' to minimal in cases where the covariance kernel generates RKHSs that are sufficiently rich to contain many (possibly most) interesting mean functions. Revisiting another well-known covariance kernel, Section \ref{Subsection: Two-Sided Brownian Motion} provides evidence suggesting that the answer to that question will be affirmative in certain special cases.

\subsection{Bilinear Covariance Kernel}\label{Subsection: Bilinear Covariance Kernel}

Suppose $\CC(\Bs,\Bt)=\Bs'\mathbf{\Sigma}\Bt$ for some (symmetric and) positive definite $\mathbf{\Sigma}$; that is, suppose $\CC$ is a bilinear form. Then $\CG^{\mu}(\Bs)=\Bs'\mathbf{\dot{\CG}^{\mu}}$, where $\mathbf{\dot{\CG}^{\mu}}=(\CG^{\mu}(\Be_1),\dots,\CG^{\mu}(\Be_d))'\thicksim \CN(\mathbf{0},\mathbf{\Sigma})$. If also $\mu$ is a quadratic form $\mu(\Bs)=-\Bs'\mathbf{\Gamma}\Bs/2$ (for some symmetric and positive definite $\mathbf{\Gamma}$), then
\begin{equation*}
    \hat{\Bs}=\argmax_{\Bs\in\R^d}\left\{-\frac{1}{2}\Bs'\mathbf{\Gamma}\Bs+\Bs'\mathbf{\dot{\CG}^{\mu}}\right\}=\mathbf{\Gamma}^{-1}\mathbf{\dot{\CG}^{\mu}}
    \thicksim\CN(\mathbf{0},\mathbf{\Gamma}^{-1}\mathbf{\Sigma}\mathbf{\Gamma}^{-1}).
\end{equation*}
Thus, this case covers asymptotically normal estimators by writing the normal random limit as the argmax of a Gaussian process. 
More generally, under regularity conditions including invertibility of the gradient $\mathbf{\dot{\mu}}$ of $\mu$, we have $\hat{\Bs}=\mathbf{\dot{\mu}}^{-1}(-\mathbf{\dot{\CG}^{\mu}})$, implying in turn that the distributional properties of $\hat{\Bs}$ can be deduced with the help of standard tools.

In other words, if $\CC$ is bilinear, then conditions for continuity of $F_{\hat{\Bs}}$ can be formulated without invoking the results of this paper. In fact, it turns out that Theorem \ref{Theorem} is completely silent about the case where $\CC$ is bilinear because in that case $\mu$ satisfies Assumption \ref{Assumption}\textit{\ref{Assumption: RKHS}} if and only if it is a linear form $\mu(\Bs)=\Bs'\mathbf{\dot{\mathbf{\mu}}}$ (for some $\dot{\mathbf{\mu}}\in\R^d$) , in which case Assumption \ref{Assumption}\textit{\ref{Assumption: Existence of argmax}} fails because $\CG(\Bs)=\Bs'(\mathbf{\dot{\mathbf{\mu}}}+\mathbf{\dot{\mathbf{\CG}}^{\mu}})$ does not admit a maximizer over $\R^d$. To summarize, our results complement existing techniques, Assumption \ref{Assumption}\textit{\ref{Assumption: RKHS}} being very restrictive precisely when the covariance kernel of $\CG$ is so simple that no new methods are needed in order to analyze the distribution of $\hat{\Bs}$.

\subsection{Two-Sided Brownian Motion}\label{Subsection: Two-Sided Brownian Motion}

Our motivating examples have the common feature that if $d=1$, then $\CC$ is proportional to $\CC_{\mathtt{BM}}$. More generally, the examples have the feature that for any $d$, $\CC$ is a (linear) functional of $\CC_{\mathtt{BM}}$, a feature which in turn would appear to be shared by most other examples of estimators satisfying \eqref{equation: Chernoff-type asymptotics} with a covariance kernel that is not bilinear. It is therefore of interest to further investigate the continuity properties of $F_{\hat{\Bs}}$ in the special case where $\CG^{\mu}$ is a two-sided Brownian motion.

Accordingly, suppose $d=1$ and suppose $\CC=\sigma^2\CC_{\mathtt{BM}}$ for some $\sigma^2>0$. Also in this case Assumption \ref{Assumption}\textit{\ref{Assumption: RKHS}} reduces to a primitive condition on $\mu$. Indeed, proceeding as in \citet[Example 2.6.7]{Gine-Nickl_2016_Book} it can be shown that Assumption \ref{Assumption}\textit{\ref{Assumption: RKHS}} holds if and only if $\mu$ is zero-preserving and absolutely continuous with a locally square integrable (weak) derivative.

In the leading special case where
\begin{equation}\label{equation: power mean}
    \mu(s)=-c|s|^{\gamma}\qquad\text{for some }c,\gamma>0,
\end{equation}
Theorem \ref{Theorem} therefore implies that $F_{\hat{\Bs}}$ is continuous whenever $\gamma>1/2$. The same condition on $\gamma$ is necessary and sufficient in order to deduce continuity of $F_{\hat{\Bs}}$ by applying \citet[Lemma A.2]{Cattaneo-Jansson-Nagasawa_2024_AoS}, which replaces Assumption \ref{Assumption}\textit{\ref{Assumption: RKHS}} with the Brownian motion-specific assumption
\begin{equation}\label{equation: CCJ mean condition}
    \lim_{\eta\downarrow0}\frac{\mu(s+\eta)-\mu(s)}{\sqrt{\eta}}=0 \qquad \text{for every }s\in\R.
\end{equation}

Maintaining the assumption that $\CG^{\mu}$ is a two-sided Brownian motion, but looking beyond mean functions of the form \eqref{equation: power mean}, the assumption \eqref{equation: CCJ mean condition} is slightly more general than Assumption \ref{Assumption}\textit{\ref{Assumption: RKHS}}. To see this, notice on the one hand that if $\mu$ is absolutely continuous with (weak) derivative $\dot{\mu}$, then
\begin{equation*}
    \left|\frac{\mu(s+\eta)-\mu(s)}{\sqrt{\eta}}\right|=\left|\frac{\int_{s}^{s+\eta}\dot{\mu}(t)\text{d}t}{\sqrt{\eta}}\right|\leq\sqrt{\int_{s}^{s+\eta}\dot{\mu}(t)^2\text{d}t}
\end{equation*}
by the Cauchy-Schwarz inequality, so \eqref{equation: CCJ mean condition} holds if $\dot{\mu}$ is locally square integrable. On the other hand, the function
\begin{equation*}
    s \mapsto -|s|\sin\left(\frac{\pi/2}{\min\{1,|s|\}}\right)
\end{equation*}
satisfies \eqref{equation: CCJ mean condition}, but is not absolutely continuous (on intervals containing zero).

In other words, in the special case of a two-sided Brownian motion, Assumption \ref{Assumption}\textit{\ref{Assumption: RKHS}} can be replaced by a slightly weaker assumption, namely \eqref{equation: CCJ mean condition}, which can accommodate departures from absolute continuity. What is less clear, but arguably more interesting, is whether Assumption \ref{Assumption}\textit{\ref{Assumption: RKHS}} imposes unduly restrictive constraints on $\gamma$ also when $\mu$ is of the form \eqref{equation: power mean} near zero. In the remainder of this section, we attempt to shed light on that question.

When $\mu$ is of the form \eqref{equation: power mean}, the condition $\gamma>1/2$ serves dual purposes when verifying the fact that $F_{\hat{\Bs}}$ is continuous. On the one hand, because \eqref{equation: argmax existence, covariance} holds with $H=1/2$, the condition $\gamma>1/2$ ensures that the tail behavior of $\mu$ is such that \eqref{equation: argmax existence, mean} holds. In addition, $\gamma>1/2$ is necessary to ensure that $\mu$ is sufficiently well behaved near zero that the weak derivative of $\mu$ is locally square integrable. To shed more light on Assumption \ref{Assumption}\textit{\ref{Assumption: RKHS}}, we disentangle the dual implications of $\gamma>1/2$ in \eqref{equation: power mean} by considering the mean function
\begin{equation}\label{equation: power mean, piecewise}
    \mu(s)=-c|s|\min\{1,|s|\}^{\gamma-1}\qquad\text{for some }c,\gamma>0,
\end{equation}
which automatically satisfies \eqref{equation: argmax existence, mean}, but satisfies Assumption \ref{Assumption}\textit{\ref{Assumption: RKHS}} only when $\gamma>1/2$. The following example shows that every $\gamma<1/2$ admits a $c=c(\gamma,\sigma^2)>0$ such that if $\mu$ is given by \eqref{equation: power mean, piecewise}, then $F_{\hat{\Bs}}$ is discontinuous, suggesting in turn that for this canonical covariance kernel at least, Assumption \ref{Assumption}\textit{\ref{Assumption: RKHS}} is close to minimal in Theorem \ref{Theorem}.

\begin{examp}
    Fix $\gamma\in(0,1/2)$ and note that with probability one the sample paths of $\CG^{\mu}$ are $\gamma$-H\"older continuous on $[0,1]$ \citep[e.g., Theorem 14.5 (iii) of][]{Kallenberg_2021_Book}. As a consequence, there exists a constant $c=c(\gamma,\sigma^2)$ such that
    \begin{equation}\label{equation: Choice of c}
        \P\left[\sup_{s\in(0,1]} s^{-\gamma}\CG^{\mu}(s)\geq c\right]<1/4.
    \end{equation}
    Fixing any such $c$, let $\CG=\CG^{\mu}+\mu$, where $\mu$ is defined as in \eqref{equation: power mean, piecewise}.
    
    By \eqref{equation: Choice of c}, we have
    \begin{equation*}
        \P[\hat{\Bs}\in(0,1]] \leq \P\left[\sup_{s\in(0,1]} \CG(s)\geq 0\right] = \P\left[\sup_{s\in(0,1]} s^{-\gamma}\CG^{\mu}(s)\geq c\right]< 1/4, 
    \end{equation*}
    where the first inequality uses $\CG(0)=0$.
    Similarly,
    \begin{align*}
        \P[\hat{\Bs}\in[1,\infty)] &\leq \P\left[\sup_{s\in[1,\infty)} \CG(s)\geq 0\right] = \P\left[\sup_{s\in[1,\infty)} s^{-1}\CG^{\mu}(s) \geq c\right] \\
        &\leq \P\left[\sup_{s\in[1,\infty)} s^{\gamma-1}\CG^{\mu}(s)\geq c\right] = \P\left[\sup_{s\in(0,1]} s^{1-\gamma}\CG^{\mu}(1/s)\geq c\right] < 1/4, 
    \end{align*}
    where the first inequality uses $\CG(0)=0$, the second inequality uses $\gamma\geq0$, and the third inequality uses \eqref{equation: Choice of c} and the time inversion property of Brownian motion \citep[e.g., Lemma 14.6 (i) of][]{Kallenberg_2021_Book}. Therefore, $\P[\hat{\Bs}>0] \leq \P[\hat{\Bs}\in(0,1]] + \P[\hat{\Bs}\in[1,\infty)] < 1/2$. Likewise, $\P\left[\hat{\Bs}<0\right] < 1/2$, so $\P[\hat{\Bs}=0] = 1 - \P[\hat{\Bs}<0] - \P[\hat{\Bs}>0] > 0$, implying in particular that $F_{\hat{\Bs}}$ is discontinuous at zero.
\end{examp}

\bibliographystyle{ecta}
\bibliography{CCJN_2025_ECMA--bib}

\end{document}